\documentclass[twoside,twocolumn,9pt]{article}
\usepackage{extsizes}
\usepackage[super,sort&compress,comma]{natbib} 
\usepackage[version=3]{mhchem}
\usepackage[left=1.5cm, right=1.5cm, top=1.785cm, bottom=2.0cm]{geometry}
\usepackage{balance}
\usepackage{mathptmx}
\usepackage{sectsty}
\usepackage{graphicx} 
\usepackage{lastpage}
\usepackage[format=plain,justification=justified,singlelinecheck=false,font={stretch=1.125,small,sf},labelfont=bf,labelsep=space]{caption}
\usepackage{float}
\usepackage{fancyhdr}
\usepackage{fnpos}
\usepackage[english]{babel}
\addto{\captionsenglish}{%
  
}
\usepackage{array}
\usepackage{droidsans}
\usepackage{charter}
\usepackage[T1]{fontenc}
\usepackage[usenames,dvipsnames]{xcolor}
\usepackage{setspace}
\usepackage[compact]{titlesec}

\usepackage{siunitx}
\usepackage{amssymb}
\usepackage{comment}
\usepackage{nicefrac}
\usepackage{booktabs}
\usepackage{soul}
\usepackage{stfloats} 

\definecolor{r3}{rgb}{0.89, 0.0, 0.13}

\newcommand{\be}{\begin{equation}}
\newcommand{\ee}{\end{equation}}
\newcommand{\bea}{\begin{eqnarray}}
\newcommand{\eea}{\end{eqnarray}}
\newcommand{\Ueff}{U_\text{eff}}
\newcommand{\Btex}{B_{2,\text{ex}}}
\newcommand{\Bts}{B_2^*}
\newcommand{\Bnf}{B_\text{NF}^*}
\newcommand{\seff}{\sigma_\text{eff}}
\newcommand{\seffISO}{\sigma_\text{eff}^\text{NF,iso}}
\newcommand{\seffBINO}{\sigma_\text{eff}^\text{cp}}
\newcommand{\seffNF}{\sigma_\text{eff}^\text{NF}}
\newcommand{\seffdry}{\sigma_\text{eff}^\text{dry}}
\newcommand{\seffrep}{\sigma_\text{eff}^\text{rep}}
\newcommand{\eKF}{\epsilon_\text{KF}}
\newcommand{\bKF}{\beta_{c,\text{KF}}}

\newcommand{\refabbrv}{ref.}
\newcommand{\refabbrvplural}{refs.}
\newcommand{\sectionabbrv}{Section}
\newcommand{\figureabbrv}{Fig.}
\newcommand{\tableabbrv}{Table}
\newcommand{\equationabbrv}{eqn}

\definecolor{cream}{RGB}{222,217,201}

\begin{document}

\pagestyle{fancy}
\thispagestyle{plain}
\fancypagestyle{plain}{
\renewcommand{\headrulewidth}{0pt}
}

\makeFNbottom
\makeatletter
\renewcommand\LARGE{\@setfontsize\LARGE{15pt}{17}}
\renewcommand\Large{\@setfontsize\Large{12pt}{14}}
\renewcommand\large{\@setfontsize\large{10pt}{12}}
\renewcommand\footnotesize{\@setfontsize\footnotesize{7pt}{10}}
\makeatother

\renewcommand{\thefootnote}{\fnsymbol{footnote}}
\renewcommand\footnoterule{\vspace*{1pt}%
\color{cream}\hrule width 3.5in height 0.4pt \color{black}\vspace*{5pt}} 
\setcounter{secnumdepth}{5}

\makeatletter 
\renewcommand\@biblabel[1]{#1}            
\renewcommand\@makefntext[1]%
{\noindent\makebox[0pt][r]{\@thefnmark\,}#1}
\makeatother 
\renewcommand{\figurename}{\small{Fig.}~}
\sectionfont{\sffamily\Large}
\subsectionfont{\normalsize}
\subsubsectionfont{\bf}
\setstretch{1.125} 
\setlength{\skip\footins}{0.8cm}
\setlength{\footnotesep}{0.25cm}
\setlength{\jot}{10pt}
\titlespacing*{\section}{0pt}{4pt}{4pt}
\titlespacing*{\subsection}{0pt}{15pt}{1pt}

\fancyfoot{}
\fancyfoot[RO]{\footnotesize{\sffamily{\thepage}}}
\fancyfoot[LE]{\footnotesize{\sffamily{\thepage}}}

\fancyhead{}
\renewcommand{\headrulewidth}{0pt} 
\renewcommand{\footrulewidth}{0pt}
\setlength{\arrayrulewidth}{1pt}
\setlength{\columnsep}{6.5mm}
\setlength\bibsep{1pt}

\makeatletter 
\newlength{\figrulesep} 
\setlength{\figrulesep}{0.5\textfloatsep} 

\newcommand{\topfigrule}{\vspace*{-1pt}%
\noindent{\color{cream}\rule[-\figrulesep]{\columnwidth}{1.5pt}} }

\newcommand{\botfigrule}{\vspace*{-2pt}%
\noindent{\color{cream}\rule[\figrulesep]{\columnwidth}{1.5pt}} }

\newcommand{\dblfigrule}{\vspace*{-1pt}%
\noindent{\color{cream}\rule[-\figrulesep]{\textwidth}{1.5pt}} }

\makeatother

\twocolumn[
  \begin{@twocolumnfalse}
\sffamily
\begin{tabular}{m{1.5cm} p{14cm} }

 & \noindent\LARGE{\textbf{Effective patchiness from critical points of a coarse-grained protein model with explicit shape and charge anisotropy}} \\
\vspace{0.3cm} & \vspace{0.3cm} \\

 & \noindent\large{Jens Weimar, Frank Hirschmann and Martin Oettel$^{\ast}$} \\

& \noindent\normalsize{Colloidal model systems are successful in rationalizing emergent phenomena like aggregation, rheology and phase behaviour of protein solutions. Colloidal theory in conjunction with isotropic interaction models is often employed to estimate the stability of such solutions. In particular, a universal criterion for the reduced second virial coefficient at the critical point $\Bts$ is frequently invoked which is based on the behavior of short-range attractive fluids (Noro-Frenkel rule, $\Bts\approx-1.5$). However, if anisotropic models for the protein-protein interaction are considered, e.g. the Kern-Frenkel (KF) patchy particle model, the value of the $\Bts$ criterion is shifted to lower values and explicitly depends on the number of patches. If an explicit shape anisotropy is considered, as e.g. in a coarse-grained protein model, the normalization of $\Bts$ becomes ambiguous to some extent, as no unique exclusion volume can be defined anymore. Here, we investigate a low-resolution, coarse-grained model for the globular protein bovine serum albumin (BSA) and study effects of charge-anisotropy on the phase diagram (determined by simulations) at the isoelectric point. We present methods of assigning an ``effective patchiness'' to our protein model by comparing its critical properties to the KF model. We find that doubling the native charges increases the critical temperature $T_c$ by $\approx\SI{14}{\percent}$ and that our BSA model can be compared to a 3 to 5 patch KF model. Finally, we argue that applying existing $\Bts$ criteria from colloidal theory should be done with care, due to multiple, physically plausible ways of how to assign effective diameters to shape-anisotropic models. 
} \\

\end{tabular}

 \end{@twocolumnfalse} \vspace{0.6cm}

  ]

\renewcommand*\rmdefault{bch}\normalfont\upshape
\rmfamily
\section*{}
\vspace{-1cm}

\footnotetext{\textit{Institute for Applied Physics, University of T\"ubingen, Auf der Morgenstelle 10, 72076 T\"ubingen, Germany. E-mail: martin.oettel@uni-tuebingen.de}}

\section{Introduction}

Liquid-liquid phase transitions (LLPS) in protein solutions are a common phenomenon, they are related to biological function,\cite{Peng2021} but they also play an important role in the pathogenesis of some human diseases \cite{Pande2001,Zhang2017} and also affect pathways to crystallization.\cite{Wolde1997,Zhang2011} Concepts from soft matter science (liquid state and colloidal theory) have been useful to see general and universal trends in protein phase behavior.\cite{Schurtenberger2020} This has been possible despite the fact that proteins are in general very anisotropic in shape and mutual interactions. Furthermore, interactions can be specific and non-specific and may sensitively depend on solvent conditions, and additionally proteins are flexible to a highly varying degree. In the simplest approach, isotropic colloidal models (hard spheres with short-ranged attractions) are used to explain features of protein phase diagrams. Notably these are the metastability of liquid-liquid demixing with respect to crystallization and an overall criterion for the necessary strength of attraction for the onset of this demixing. The latter has been proposed in \refabbrvplural~\citenum{Vliegenthart2000} and \citenum{Noro2000} and is based on an approximately universal phase behavior of short-range attractive colloidal models. It is formulated in terms of a reduced second virial coefficient $\Bts$ at the critical temperature which takes the approximate value $\Bts \approx -1.5$ for the investigated colloidal models and has been suggested also to hold for protein solutions. In the following we refer to this finding as ``Noro-Frenkel rule'' or ``law of corresponding states''. 

A comprehensive study from the group of the late Stefan Egelhaaf \cite{Platten2015} finds a rather good agreement with the Noro-Frenkel rule if one scales the low-concentration part of the protein binodal with the corresponding part from a short-range square well fluid. On the other hand, the anisotropy of the attractive interactions between proteins frequently gives rise to clustering behavior which is reminiscent of the one seen in patchy particle models. For the latter (provided the interactions are only patchy and hard core), however, the location of the gas-liquid critical point is not consistent with the Noro-Frenkel rule and mainly the number of patches determine the reduced second virial coefficient, e.g. $\Bts\approx-30$ for three patches, $\Bts\approx-4...-5$ for four patches etc..\cite{Foffi2007} Thus one would expect also a deviation of $\Bts$ from the Noro-Frenkel value for protein solutions if there is enough anisotropy in the interactions. Real protein-protein interactions are neither purely isotropic nor purely patchy, but always a combination of both. A study about patchy square well fluids, which represent such a combination, did not observe a law of corresponding states, i.e. in the form of $\Bts$ vs. patch number.\cite{Liu2009} This implies that using universal $\Bts$ criteria should only be applicable if the interactions of the model at hand are considered either mostly isotropic or mostly patchy. 

Anisotropy is of course fully present if an atomistic protein model in explicit solvent is used.
From a simulation point of view, it is still difficult to investigate phase separation in all-atom models for biomolecules and currently is restricted to small proteins and peptides.\cite{Blanco2022} Here it is better to resort to coarse-grained (CG) models in which the protein is built from pseudo-atoms that unite multiple atoms into a single, bigger one. There exist numerous coarse-graining approaches (for reviews see e.g. \refabbrvplural~\citenum{Voth2022,Blanco2022,Kolinski2016}), with each of them operating at a certain resolution level. High-resolution CG models like the well-known MARTINI\cite{Marrink2007} force field are often used with explicit water (e.g. for intrinsically disordered proteins\cite{Fagerberg2023} or nanoparticles \cite{Monticelli2012}), whereas low-resolution models in which only a few effective beads represent the molecule interactions are modeled by implicit-solvent potentials, as e.g. of DLVO type. These low-resolution models have been widely employed in the literature, e.g. in studies on antibodies\cite{Chaudhri2013,Wang2018,Izadi2020,Shahfar2021} but also for more globular proteins.\cite{Gruenberger2013,Kozlowska2021}

Here we investigate the phase behavior of a coarse-grained model for the globular protein bovine serum albumine (BSA) consisting of 6 beads, using grand canonical Monte Carlo (GCMC) simulations. Interactions between beads are of DLVO type, the total BSA charge is zero but there is a fixed distribution of relative bead charges with one multiplicative parameter scaling the individual bead charges. In this model, the shape anisotropy (coming from the arrangement of the beads) determines also an asymmetry of the van der Waals attractions. Additionally, the bead charges add a further charge asymmetry to these interactions. We study the influence of charge on the critical point and the bead-bead correlations and attempt to quantify an effective patchiness from the $\Bts$-value for the investigated system. 

The motivation for studying a BSA model is linked to the existence of LLPS of BSA in solutions with trivalent cations.\cite{Zhang2008} On the one hand, these LLPS have been rationalized using patchy particle models, \cite{RoosenRunge2014,Fries2020} but on the other hand, scattering studies near the critical point in these systems show only small deviations from the Noro-Frenkel rule, thereby suggesting a small influence of patchiness on the phase diagram. Furthermore, BSA is well characterized by scattering methods in terms of its shape, dynamics and diffusion properties, \cite{Gaigalas1992,Zhang2007,Bujacz2012,Ameseder2019,Beck2021} it appears to be a rather rigid molecule suitable for modelling.

The paper is structured as follows: We first introduce our coarse-grained BSA model and its intermolecular interactions in \sectionabbrv~\ref{sec:BSAmodel}, where in total six different parameter combinations are defined. Subsequently, in \sectionabbrv~\ref{Sec:GCMCsims} the determination of critical and binodal points via GCMC simulation is described. In \sectionabbrv~\ref{sec:results} the resulting phase diagrams are shown, along with the effective potentials at criticality, the second virial coefficients and effective molecule diameters, which are calculated via four different approaches. Additionally, we present bead-bead correlation functions and discuss their charge dependence. In \sectionabbrv~\ref{sec:discussion} we discuss how an ``effective patchiness'' can be deduced from the critical point of our models with paying special attention to the ambiguity in defining effective hard diameters and thus in the normalization of the second virial coefficient (needed for defining $\Bts$). Furthermore the relationship of our results to experiments is discussed, before we conclude in \sectionabbrv~\ref{sec:conclusion}.
           
\section{Model description}\label{sec:BSAmodel}
\begin{figure}[h]
    \centering
    \includegraphics[width=1\linewidth]{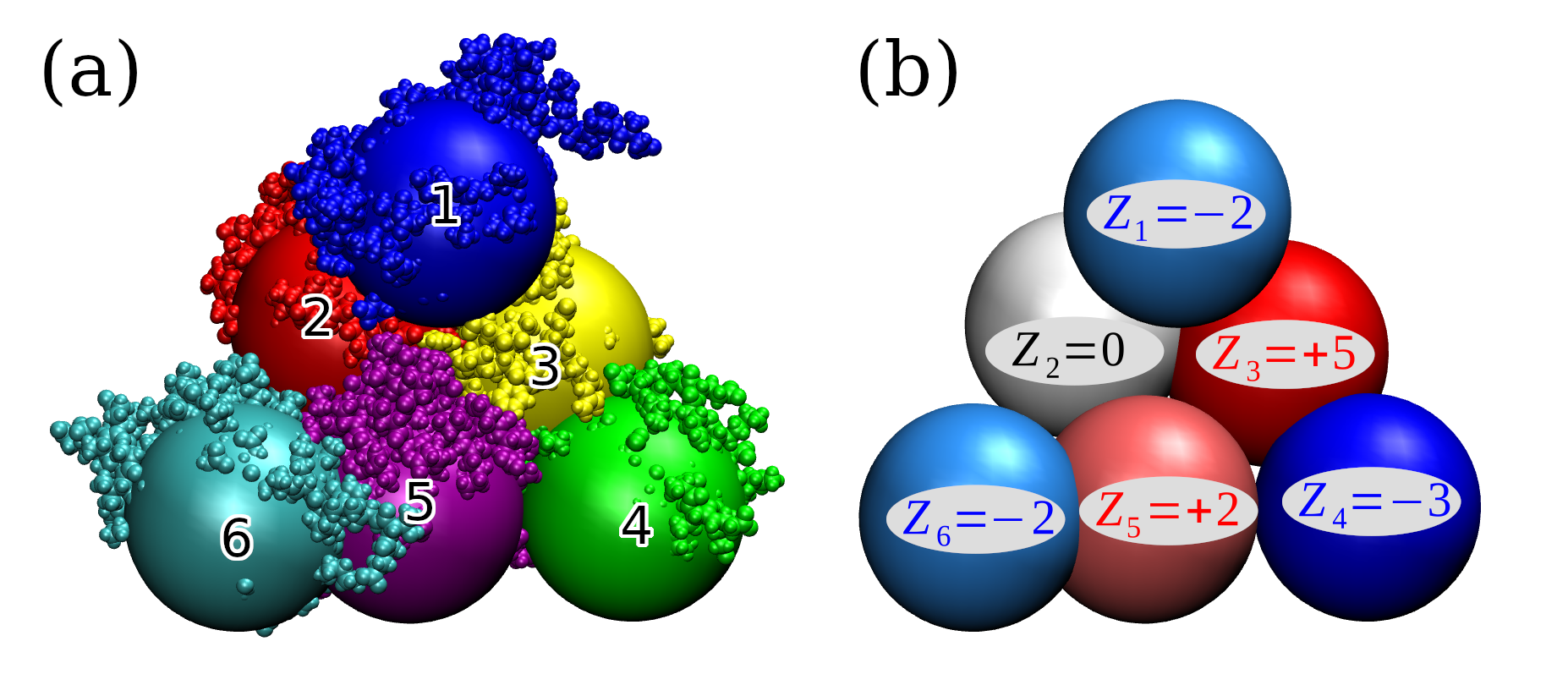}
    \caption{(a) Atomistic structure of BSA\cite{Bujacz2012} superimposed onto the domain-based, six-bead coarse-grained model of \refabbrv~\citenum{Hirschmann2023} with bead diameter $\sigma= \SI{29.52}{\angstrom}$. Missing hydrogens have been added with the pdb2gmx\cite{Abraham2015} tool. (b) Resulting coarse-grained charge distribution in units of the electron charge $e$ of BSA at the isoelectric point ($\text{pH}\approx 5.8$). Figures have been rendered using the VMD software.\cite{Humphrey1996}}
    \label{fig:BSA_molec}
\end{figure}

The BSA model investigated in this work has previously been derived and analysed via Brownian dynamics (BD) simulations in \refabbrv~\citenum{Hirschmann2023}. In brief, the authors conducted single-molecule all-atom MD simulations of a crystal-structure of BSA and subsequently used direct Boltzmann inversion to obtain coarse-grained force field parameters, yielding a flexible, six-beaded, low-resolution CG model, in which each bead represents a biological domain, see \figureabbrv~\ref{fig:BSA_molec}(a) for illustration. The study assessed the influence of flexibility in self-crowded conditions on static and dynamic solution properties like center-of-mass pair-correlations and diffusion. However, the inter-molecular interactions were taken to be purely repulsive and soft, accounting only for excluded volume of the molecules, corresponding to a high-salt regime, where electrostatic interactions are sufficiently screened.

In this work, we consider more realistic interactions between molecules by incorporating dispersion forces and screened electrostatics which give rise to gas-liquid phase transitions and coexistence. More specifically, we set the interaction potential (bead-bead potential) between two coarse-grained sites on two distinct molecules to:

\begin{equation}
\label{Eq:Utot}
    U_{\text{tot}}(r_{ij}) =  U^{}_{\text{rep}}(r_{ij}) + U^{}_{\text{vdW}}(r_{ij}) +  U^{}_{\text{el}}(r_{ij}) + U^{}_{\text{cont}}(r_{ij})
\end{equation}
where indices $i$ and $j$ refer to the $i$th and $j$th bead on the first and the second molecule (respectively) and $r_{ij}$ represents the center-to-center distance of these beads. The individual contributions are as follows.

Volume exclusions are modelled by the repulsive part of the Lennard-Jones potential $U^{}_{\text{rep}}$, which is cut-off at its minimum:
\begin{equation}\label{Eq:Urep}
U^{}_{\text{rep}}(r_{ij}) 
=
\begin{aligned}
\begin{cases}
4 \epsilon \left[	(\sigma/r_{ij})^{12}	-	(\sigma/r_{ij})^6	\right] &  \text{} \ \ r_{ij}	\leqslant	r^{\text{cut}}_{\text{rep}} \\
0 &  \text{} \ \ r_{ij}	>	r^{\text{cut}}_{\text{rep}}
\end{cases}
\end{aligned},
\end{equation}
where $\epsilon$ is the energy scale. Dimensionless energies $U^*$ and temperatures $T^*$ can be defined by $U^{*}=U/\epsilon$ and $T^{*}=k_{\text{B}}T/\epsilon$, where $k_{\text{B}}$ is the Boltzmann constant and $T$ is the temperature. The asterisks are dropped immediately and for the remainder of this work, energies and temperatures are to be understood in their respective reduced units. The cutoff distance is $r^{\text{cut}}_{\text{rep}}= {2^{1/6}}\,{\sigma} $ and the  bead  diameter $\sigma= \SI{29.52}{\angstrom}$ has been derived in \refabbrv~\citenum{Hirschmann2023}.

Dispersion interactions are modelled by the van der Waals potential $U^{}_{\text{vdW}}$ for two equally sized spheres with:\cite{Hamaker1937} 
\begin{equation}\label{Eq:UvdW}
U^{}_{\text{vdW}}(r_{ij}) 
=
\begin{aligned} 
\begin{cases}
0  &   \text{} \ \ r_{ij}	<	r^{\text{cut}}_{\text{rep}}  \\
- \frac{A_H}{12} \left( \frac{\sigma^2}{r_{ij}^2 -\sigma^2}  +\frac{\sigma^2}{r_{ij}^2} + 2 \ln \frac{r_{ij}^2-\sigma^2}{r_{ij}^2}  \right)  &   \text{} \ \ r^{\text{cut}}_{\text{rep}} \leqslant	 r_{ij}	\leqslant	r^{\text{cut}}_{\text{vdW}} \\
0  &   \text{} \ \ r_{ij}	>	r^{\text{cut}}_{\text{vdW}} 
\end{cases}
\end{aligned},
\end{equation}
where $A_H$ is the Hamaker constant (in units of $\epsilon$) and the cutoff distance has been set to $r^{\text{cut}}_{\text{vdW}} = \SI{2.5}{\sigma}$. 

Electrostatic interactions are approximated via a screened Yukawa potential $U^{}_{\text{el}}$:
\begin{equation}
\label{Eq:Uel}
    U_{\text{el}}(r_{ij}) = 
\begin{aligned} 
\begin{cases}
    k_{\text{B}}T \lambda_{\text{B}}  Z_iZ_j  \left( \frac{\exp{ \left( \kappa\sigma /2  \right)  }}{1+\kappa \sigma/2} \right)^2  \frac{\exp{ \left( -\kappa r_{ij}  \right)  }}{r_{ij}} & \text{} \ \ r_{ij}	\leqslant	r^{\text{cut}}_{\text{el}} \\
    0 & \text{} \ \ r_{ij}	>	r^{\text{cut}}_{\text{el}} \\
\end{cases}
\end{aligned},
\end{equation}
where $\lambda_{\text{B}} $ is the Bjerrum length, $Z_i$ and $Z_j$ are the electric charges on bead $i$ and $j$ (respectively, in units of the elementary charge $e$) and $\kappa^{-1}$ is the Debye screening length. In order to have a noticeable influence of charges, we choose a moderate screening length of $\kappa^{-1}=\sigma$, which corresponds to a monovalent salt concentration of $c\approx \SI{11}{\milli\mole\per\liter}$. Moreover, for simplicity, we ignore the temperature dependence of the Bjerrum length and set it to the textbook value for water at room temperature, $\lambda_{\text{B}} \approx \SI{0.7}{\nm} $. Furthermore, $r^{\text{cut}}_{\text{el}} = r^{\text{cut}}_{\text{vdW}} = \SI{2.5}{\sigma}$. The charge distribution (i.e. $Z_i$ in \equationabbrv~(\ref{Eq:Uel})) of all coarse-grained molecules within simulations is identical and determined through processing the atomistic structure of BSA (PDB ID: 4F5S, reported in \refabbrv~\citenum{Bujacz2012}) with the APBS webserver,\cite{Jurrus2018} employing the PROPKA model\cite{Sondergaard2011} to calculate the titration states of amino acids at the isoelectric point at $\text{pH}\approx 5.8$ (experiments determine a slightly deviating value of $\text{pH}\approx 4.7$) \cite{Salis2011}. Subsequently, the partial charges of all amino acid residues belonging to the same coarse-grained site are summed up to calculate the total charge of the bead, see \figureabbrv~\ref{fig:BSA_molec}(b) for the resulting charge distribution. We deliberately chose our molecules to have an overall neutral net charge, since in experiments for various globular proteins in solution with multivalent ions, aggregation of molecules and liquid-liquid phase separation (LLPS) is observed around the isoelectric point.\cite{Zhang2010,Matsarskaia2018} Additionally, in order to artificially alter the charge anisotropy of the CG molecules, we introduce a charge multiplier (CM), which uniformly scales all bead charges up or down according to: 
\begin{equation}
    Z \rightarrow  \text{CM} \, Z,
\end{equation}
such that $\text{CM}=0$ would result in molecules with no electric charges on all beads and $\text{CM}=2$ would result in a model where all charges depicted in \figureabbrv~\ref{fig:BSA_molec}(b) are doubled.

The last term $U^{}_{\text{cont}}$ in \equationabbrv~(\ref{Eq:Utot}) compensates the usage of cutoffs by making the total potential continuous. It reads: 
\begin{equation}
U^{}_{\text{cont}}(r_{ij}) = 
\begin{aligned} 
\begin{cases}
\vspace{0.1cm}
\!\begin{aligned}
&- U^{}_{\text{rep}}(r^{\text{cut}}_{\text{rep}}) - U^{}_{\text{vdW}}(r^{\text{cut}}_{\text{vdW}})  \\
& \ \ - U^{}_{\text{el}}(r^{\text{cut}}_{\text{el}}) + U^{}_{\text{vdW}}(r^{\text{cut}}_{\text{rep}})
\end{aligned}
& \text{} \ \ r_{ij}	<	r^{\text{cut}}_{\text{rep}} 
\\
\vspace{0.1cm}
 - U^{}_{\text{vdW}}(r^{\text{cut}}_{\text{vdW}})   - U^{}_{\text{el}}(r^{\text{cut}}_{\text{el}})  
& \text{} \ \ r^{\text{cut}}_{\text{rep}} \leqslant	 r_{ij}	\leqslant	r^{\text{cut}}_{\text{vdW}} 
\\ 
0 & \text{} \ \ r_{ij}	>	r^{\text{cut}}_{\text{vdW}} = r^{\text{cut}}_{\text{el}} 
\\
\end{cases}
\end{aligned}.
\end{equation}

\begin{figure}[h]
    \centering
    \includegraphics[width=0.8\linewidth]{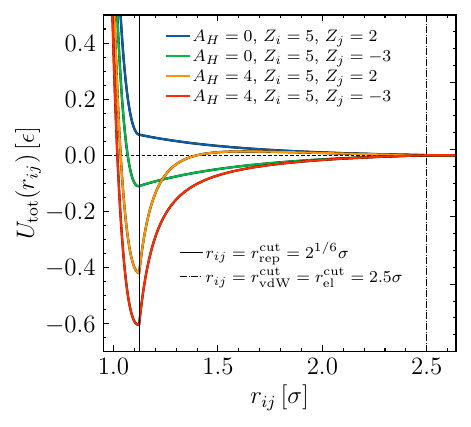}
    \caption{Total bead-bead potential according to \equationabbrv~(\ref{Eq:Utot}) for typical parameters.
    }
    \label{fig:pot_plot}
\end{figure}

For illustration, the complete interaction potential specified by \equationabbrv~(\ref{Eq:Utot}) is depicted in \figureabbrv~\ref{fig:pot_plot} for some typical parameters.

In our simulations the CG molecules are rigid bodies without internal degrees of freedom. Relative bead positions are fixed, corresponding to the average structure present in coarse-grained BD simulations at infinite dilution, as described in \refabbrv~\citenum{Hirschmann2023}. In this study, we investigate different parameter compositions, i.e. the charge anisotropy is varied by choosing $\text{CM} \in [0,1,2]$ and it is either $A_H=4$ or $T=1$ fixed (while keeping either $T$ or $A_H$ variable). Overall, this yields six different systems for which phase behavior can be studied, see \tableabbrv~\ref{tab:resultsanisotropic} for an overview. In the simulations of the $A_H=\text{const.}$-systems a physical critical temperature is determined. In contrast, in the $T=\text{const.}$-systems one determines the needed bead-bead attraction strength for criticality at ambient temperatures. 

The value of $A_H=4$ appears to be a reasonable estimate for the bead-bead attraction strength for BSA solutions with monovalent salt, since an experimental structure factor at $\text{pH}\approx7.0$, $c_{\text{BSA}}=\SI{174}{\milli\gram\per\milli\liter}$ and $c_{\text{NaCl}}=\SI{18}{\milli\mole\per\liter}$ at room temperature\cite{FajunPC2024} is roughly similar to the center-of-mass structure factor from our simulations at equivalent conditions (the methodology for obtaining experimental structure factors is described in \refabbrv~\citenum{Zhang2007}). On the other hand in the experiments with BSA solutions with trivalent salts the attraction strength between molecules is a tunable parameter at constant temperature,\cite{RoosenRunge2014} in our model this would be equivalent to a variable bead-bead attraction strength $A_H$ and a variable $\text{CM}$.

\begin{table*}[h]
	\centering 
	\caption{Summary of investigated system parameters and results for the critical Hamaker constants $A_{H,c}$, temperatures $T_c$, densities $\rho_c$ (in units of ${\sigma}^{-3}$) and $B_2$ values (in units of ${\sigma}^{3}$). Also shown are critical temperatures $T_c^{\text{iso}}$, densities $\rho_c^{\text{iso}} $ and $B_2$ values of the effective, isotropic systems (see \sectionabbrv~\ref{sec:res_ueff}). We estimate the relative statistical error for $T_c$ and $A_{H,c}$ to be $ \pm \, \SI{0.1}{\percent}$ and for $\rho_c$ $\pm \, \SI{0.4}{\percent}$. The relative total error of $B_2$ is $\pm \,\SI{0.7}{\percent}$  }
        \begin{tabular}{@{\extracolsep{\fill}}llllllllllll}
		\hline
            System & CM & $A_H$ & $T$ & $A_{H,c}$ & $T_c$ &  $\rho_c$ & $\nicefrac{A_{H,c}}{k_B \, T_c}$ & $B_2(T_c)$ & $T_c^{\text{iso}}$ & $\rho_c^{\text{iso}} $ & $B_2(T_c^{\text{iso}})$ \\
		\hline
  		I & 0 & 4 & variable  & 4 & 0.405 & 0.059 & 9.88 & -28.9 & 1.12 & 0.18 & -6.41 \\ 
		II & 0 & variable  & 1 & 8.59 & 1 & 0.061 &8.59 & -27.5 & 2.95 & 0.21 & -5.63 \\
		III & 1 & 4 & variable & 4 & 0.411 & 0.059 &9.74 & -28.6 & 1.02 & 0.17 & -7.29 \\ 
		IV & 1 & variable & 1 & 8.46 & 1 & 0.061 &8.46 & -27.4 & 2.61 & 0.19 &  -6.55 \\
		V & 2 & 4 & variable & 4 & 0.470 & 0.054 &8.51 &  -35.6 & 1.00 & 0.12 &  -10.7 \\ 
		VI & 2 & variable & 1  & 7.44 & 1  & 0.054 &7.44 & -35.7 & 2.25 & 0.13 & -9.9 \\
		\hline
	\end{tabular}
	\label{tab:resultsanisotropic}
\end{table*}

\section{Simulation methods}\label{Sec:GCMCsims}

The CG molecules were simulated in a grand canonical Monte Carlo scheme with textbook \cite{sciortinobook} insertion, deletion and roto-translation moves (with $5\,\%$, $5\,\%$ and $\,90\%$ attempt probability, respectively). Orientations of molecules were represented via quaternions.\cite{animationwithquaternions} The simulations were performed inside a cubic simulation box of side length $L=10 \, \sigma$. The main goal of our calculations is to obtain the critical points of the BSA systems. We additionally compute some binodal points to estimate the shape of the phase diagram. In this respect a finite size study was not conducted, since the expected finite size effects on the critical points are estimated not to be significant for the purpose of the present work.\cite{sciortinobook}  However, we expect that our binodal points are affected by finite size effects to some extend, especially those close to the critical points. 

For obtaining the binodal points in the finite system close to $T_c$, histograms of the instantaneous molecule number $N$ in the simulation box at different temperatures and values for the chemical potential $\mu$ were recorded and scanned for the appearance of double peaks. At the coexistence value for the chemical potential the area below these peaks must be equal. In order to obtain equal areas a fitting and optimization process using histogram reweighting\cite{reweighting1,reweighting2,reweighting3} was employed. Reweighting towards equal area gave new values for the chemical potential at which new histograms were simulated, whose peaks are possibly modified compared to the reweighted histogram. The procedure was repeated until the area difference below the peaks was below a chosen limit. The final peak positions give the average number of molecules in the coexisting gas and liquid states from which the corresponding number densities $\rho=N/L^3$ are determined (binodal points).
\begin{figure}[h]
    \centering
    \includegraphics[width=0.5\textwidth]{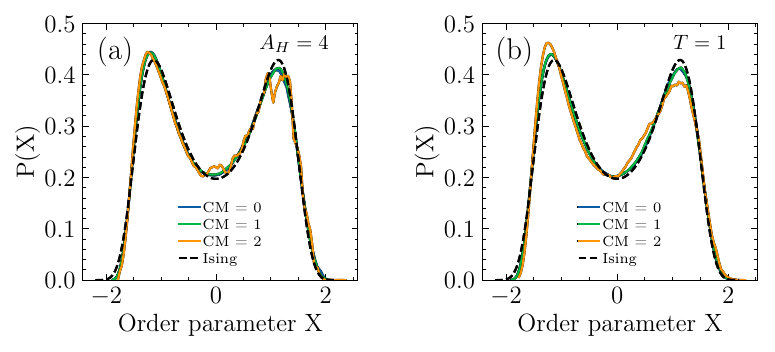}   
    \caption{Fits of the order parameter $X$ on the universal Ising curve for all six different subsystems. $X$ connects particle number and energy through a mixing parameter $s$ as $X = N - s \, E$ and was scaled on the Ising curve (zero mean, unitary variance, cf. \refabbrvplural~\citenum{sciortinobook,wilding}). (a) shows fits for systems with constant $A_H=4$ (with the critical temperature $T_c$ as fit parameter), and (b) shows fits for the systems with a constant temperature $T=1$ (with the critical value of the Hamaker constant $A_{H,c}$ as fit parameter).}
    \label{fig:uniisinganisofits}
\end{figure}

Critical points of the different BSA systems (listed in \tableabbrv~\ref{tab:resultsanisotropic}) were determined by matching their double-peak histograms to the universal Ising distribution, following the methodology of \refabbrvplural~\citenum{sciortinobook} and \citenum{wilding}. In short, near the critical temperature and the critical chemical potential, where the histograms already looked similar to the universal Ising distribution, more simulations with slightly adjusted $\mu$ and $T$ were performed. Second, by performing histogram reweighting on this data for new values $\mu'$, $T'$ and the mixing parameter $s$ (which connects the system energy $E$ and $N$ to the Ising-like order parameter $X=N-sE$ ) as fit parameters, we determine the critical temperature $T_c$ as the value of $T'$ which gives the best fit result on the universal Ising distribution. In systems where the critical Hamaker constant was determined, the simulation and reweighting process was performed with different values of $A_H$ (estimated by searching for equal area in the histograms) until the fitted value for the critical temperature $T_c$ was very close to unity and then $A_{H,c}$ was computed as $A_{H,c} = \nicefrac{A_H }{T_c}$ (the correction due to $T_c \neq 1$ was always below $0.2\si{\percent}$.)
\figureabbrv~\ref{fig:uniisinganisofits} shows fits on the universal Ising curve for the six different parameter combinations. The resulting critical densities $\rho_c$, temperatures $T_c$ and Hamaker constants $A_{H,c}$ are listed in \tableabbrv~\ref{tab:resultsanisotropic}. 
Bead-bead correlations (see \sectionabbrv~\ref{sec:res_beadcorr}) were recorded in a larger box with the same GCMC code but with insertion and deletion moves disabled, which corresponds to a canonical $NVT$ ensemble.

For obtaining the critical points $\numrange{50}{60}$ independent systems have been simulated, each of them with $10^9$ GCMC moves in total. For the binodal points and the bead-bead correlations 10 resp. 80 independent systems have been employed, with $\numrange{1e9}{2e9}$ and $10^9$ GCMC moves, respectively. We also note that obtaining binodal points proved difficult, especially for low temperatures, due to systems remaining in one phase even for long simulation runs. To this end, more elaborate simulation techniques could be applied, e.g. successive umbrella sampling,\cite{Virnau2004} which was out of the scope of this study.

\section{Results}\label{sec:results}
\subsection{Phase diagrams and critical points}
\begin{figure}[H]
    \centering
    \includegraphics[width=0.5\textwidth]{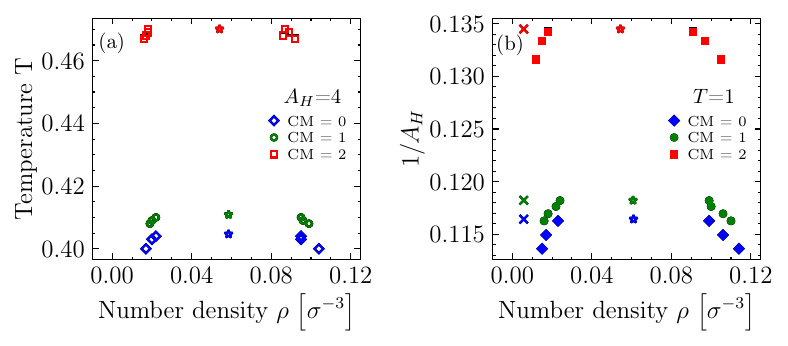}
    \caption{Binodals of differently charged systems (see \tableabbrv~\ref{tab:resultsanisotropic}) with fixed Hamaker constant $A_H=4$ (a) and fixed temperature $T=1$ (b). Critical points are indicated by colored stars and state points where bead-bead correlation functions $g_{ij}(r_{ij})$ have been recorded are indicated by colored crosses (see \figureabbrv~\ref{fig:beadcorrelations} in \sectionabbrv~\ref{sec:res_beadcorr}).}
    \label{fig:phasediagram}
\end{figure}

Our results for the binodals of the three differently charged systems with $\text{CM}=0,1,2$ and $A_H=4$, as well as the three systems with $T=1$ (see \tableabbrv~\ref{tab:resultsanisotropic}) are depicted in \figureabbrv~\ref{fig:phasediagram}. 
In general, the resulting phase diagrams obtained for systems with either variable $T$ or variable $A_H$ are in shape very similar to each other, but not exactly identical, since even for the completely uncharged systems the Hamaker constant with our definition is not just an inverse temperature. Starting from the uncharged molecules ($\text{CM}=0$) and adding the native charges ($\text{CM}=1$) to our molecules introduces additional attractive interactions to the systems and shifts $T_c$ upwards by (only) $\approx \SI{1.5}{\percent}$ ($A_{H,c}$ downwards by $\approx \SI{1.5}{\percent}$), and also inflicts minute changes to the slope of the coexistence curves. However, more pronounced effects are seen if one changes $\text{CM}$ from 1 to 2, as this shifts $T_c$ upwards by another $\approx \SI{14}{\percent}$ and $A_{H,c}$ downwards by $\approx \SI{12}{\percent}$. We attribute these additional attractions to the growing, leading order, electrical dipole moment of our molecules, and note that this physical quantity has been observed to play an important parameter for loci of coexistence regions in protein-like fluids.\cite{Blanco2016} Overall, our findings indicate that the stability in our BSA systems for $\text{CM}=1$ is mainly governed by the van der Waals interactions and not so much by electrostatics.

From the results in \tableabbrv~\ref{tab:resultsanisotropic} we infer that the effective temperature-like variable $k_BT/A_H$ is not even approximately the same at the critical point for the different systems ($A_H=\text{const.}$ and $T=\text{const.}$) at the same $\text{CM}$. The two types of systems cannot be mapped onto each other due to the soft repulsions. However the dependence of $k_BT/A_H$ on $\text{CM}$ is similar, and indeed it turns out that only $\text{CM}$ is a relevant variable (see \sectionabbrv~\ref{sec:res_sigmas}).

\subsection{Effective potentials at the critical points}\label{sec:res_ueff}

An isotropic effective potential $\Ueff(r)$ between two molecules can be defined in the usual way by the angular average:\cite{DhontBook1996}
\be 
 \label{eq:Ueff}
  \exp(-\beta \Ueff(r) ) = \int \mathrm{d} \Omega_1 \int \mathrm{d} \Omega_2 \exp( - \beta U(r,\Omega_1, \Omega_2) ),\,
\ee 
where $r$ is the center-of-mass distance between two coarse-grained molecules, $\Omega_i$ specifies the orientation of molecule $i$ (e.g. by three Euler angles), $U(r,\Omega_1, \Omega_2)$ is the total orientation-dependent pair potential between two molecules (in our case $U=\sum_{i}^6\sum_{j}^6 U_{\text{tot}}$) and the integral over orientations is normalized ($\int \mathrm{d} \Omega_i=1$). If $U$ does not depend on orientations, then $U=\Ueff$ of course. We calculate the six–dimensional integral via two methods: First, we employ Monte Carlo sampling with $10^{8}$ uniformly distributed orientations per distance $r$ with a spatial resolution of $\Delta r = 0.01\,\sigma$. Additionally, we compute \equationabbrv~(\ref{eq:Ueff}) via Gauss–Legendre quadrature and 20 grid points per angle. The obtained $\Ueff$ show negligible differences, the
resulting $B_2$ values differ on average by $\approx \SI{0.22}{\percent} $ for the two methods. For the Gauss-Legendre method using 15 grid points instead of 20 yields a relative error of obtained $B_2$ values of $\approx \SI{0.5}{\percent} $.

Through examination of molecule-molecule configurations in \equationabbrv~(\ref{eq:Ueff}) the lowest-energy state of two molecules can be determined: For the $\text{CM}=0,1$ systems this is a very compact state at $r=1.2$, which maximizes bead contacts. For $\text{CM}=2$ it is a staggered state at $r=1.4$, in which one molecule sits on top of the other but slightly rotated, such that bead 4 is near bead 3. These states, however, do not contribute to $\Ueff$ in a noticeable manner. Consistent with that is the absence of a noticeable amount of compact dimers in simulation snapshots. In these, dimers are on average further apart with varying bonding configurations.

\begin{figure}[h!]
    \centering
    \includegraphics[width=1\linewidth]{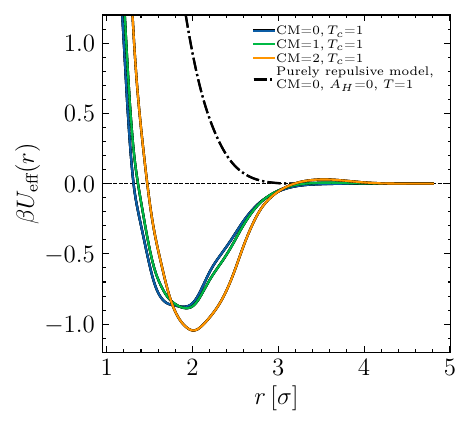}
    \caption{Effective potential $\beta U_{\text{eff}}(r)$ (\equationabbrv~(\ref{eq:Ueff})) obtained via MC integration for the critical systems with $T_c=1$ (colored) and for a purely repulsive model for which $A_H=\text{CM}=0$ and $T=1$ (black, dashed line). Effective potentials of the critical system with $A_{H,c}=4$ (not shown) are very similar to the ones depicted with the same $\text{CM}$.}
    \label{fig:ueff}
\end{figure}

We utilize the obtained effective potentials in additional GCMC simulations, in which isotropic particles interact via the six critical $\Ueff(r)$ obtained by the previously described procedure. Subsequently, via the same protocol as described in \sectionabbrv~\ref{Sec:GCMCsims} we determine critical temperatures $T_c^{\text{iso}}$ for these systems and the obtained values are listed in \tableabbrv~\ref{tab:resultsanisotropic}. In the following sections of this paper these systems are referred to as ``isotropic systems''. The obtained $T_c^{\text{iso}}$ are approximately two to threefold higher as the corresponding $T_c$ of the anisotropic models, pointing to a significant over-representation of attractions and/or under-representation of repulsions in $\Ueff(r)$. An example are higher-order multi-body interactions not being present in $\Ueff(r)$: It could be favorable for two CG molecules to form a low-energy dimer, which enters the effective potential. In the isotropic system three spheres can easily arrange to a trimer, with a threefold binding energy of the dimer, whereas three CG molecules can not easily arrange to a trimer with an equally low binding energy due to steric incompatibilities.       

\subsection{Second virial coefficients and effective molecule diameters at the critical points}\label{sec:res_sigmas}

\begin{figure}[h]
    \centering
    \includegraphics[width=0.85\linewidth]{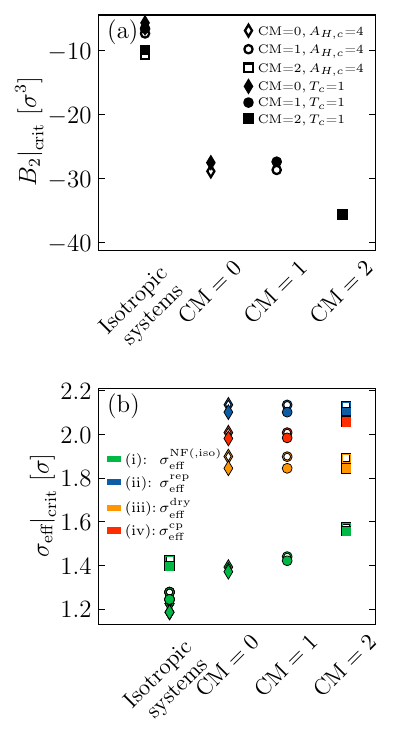}
    \caption{(a) $B_2$ values according to \equationabbrv~(\ref{eq:B2}) of all critical BSA models presented in \tableabbrv~\ref{tab:resultsanisotropic}, together with $B_2$ values of the corresponding critical isotropic systems. 
    (b) Effective molecule diameters $\sigma_{\text{eff}}$ at the critical points, calculated via the four different approaches (i)-(iv) discussed in the main text. 
    }
    \label{fig:B_sig_combo}
\end{figure}

Using the effective potential $\Ueff(r)$ of \equationabbrv~(\ref{eq:Ueff}), the second virial coefficient can be written as:
\be
 \label{eq:B2}
 B_2 = 2 \pi \int_0^\infty (  1 - \exp(-\beta \Ueff(r) ) r^2 \mathrm{d}r \;.
\ee 
The obtained values at the respective critical points of all models are depicted in \figureabbrv~\ref{fig:B_sig_combo}(a) and listed in \tableabbrv~\ref{tab:resultsanisotropic}. Increasing $\text{CM}$ from 1 to 2 introduces additional attractions between molecules, as $B_2$ noticeably decreases, whereas almost no influence is visible if comparing the uncharged molecule to the native $\text{CM}=1$ state. Within anisotropic systems sharing the same $\text{CM}$ no significant difference is visible. The values for the isotropic systems on the other hand show a stronger charge dependence and values of systems with $\text{CM}=0,1$ are more separated to systems with $\text{CM}=2$. Note that the reason for the $B_2$ values of the isotropic systems being higher than for the anisotropic ones, is that \equationabbrv~(\ref{eq:B2}) is evaluated at $T=T_c^{\text{iso}}$ and that $T_c^{\text{iso}}/T_c\approx2...3$, as pointed out before. 

For obtaining reduced second virial coefficients $\Bts=B_{2,\text{crit}}/\Btex$ it is necessary to determine an exclusion volume which represents steric incompatibilities, i.e. an effective spherical diameter $\seff$ of the molecule in terms of which $\Btex=(2\pi/3)\seff^3$. Here, we present four approaches with a certain theoretical preference for the first one and three additional methods for comparison, which however also appear natural from a physical point of view. 

(i): $\seffNF$ and $\seffISO$. In the original Noro-Frenkel (NF) analysis of various short-range attractive, isotropic systems \cite{Noro2000} $\seffNF$ is computed from the repulsive part ($v_\text{rep}(r)$) of the isotropic pair potential (determined by the WCA rule, with the pair potential $v(r)$ cut-off at the minimum and shifted such that the minimum is zero) via the Barker-Henderson (BH) rule $\seffNF = \int_0^\infty  \mathrm{d}r (1- \exp(-\beta v_\text{rep}(r))$.\cite{Barker1967} In our case, we use the effective potential for the anisotropic systems, i.e. $v(r) \to \Ueff(r)$. For the isotropic systems $\seffISO$ is calculated exactly the same way, only with $T=T_c^{\text{iso}}$ (see \figureabbrv~\ref{fig:B_sig_combo}(b) and \tableabbrv~\ref{tab:sigma_b2} (Appendix) for resulting values). 

(ii): $\seffrep$. Excluded volume results from the repulsive part of the bead-bead potentials. First we define a bead-bead potential only from repulsive interactions, $U_{\text{tot}}(r_{ij}) = U_\text{rep}(r_{ij})$, as defined in \equationabbrv~(\ref{Eq:Urep}). Then $\Ueff$ is computed from \equationabbrv~(\ref{eq:Ueff}). The resulting repulsive $\Ueff$ is charge-independent (if evaluated at the same $T$) and it is shown in comparison to the full $\Ueff$ in \figureabbrv~\ref{fig:ueff} (black, dashed line). Like before, from the BH rule the diameter $\seffrep$ of an equivalent hard sphere system can be determined. The values $\seffrep \approx 2.1 \, \sigma$ (see \figureabbrv~\ref{fig:B_sig_combo}(b) and \tableabbrv~\ref{tab:sigma_b2} (Appendix)) are much larger than $\seffNF$ from the effective potential, this is also visible in the comparison of the effective potentials in \figureabbrv~\ref{fig:ueff}.  

(iii): $\seffdry$. In \refabbrv~\citenum{Hirschmann2023} the bead diameter $\sigma= \SI{29.52}{\angstrom}$ was derived by distributing the molecule volume of BSA on six equally sized spheres. The molecule volume was obtained by computing the protein specific volume with the molecular mass of BSA. Here, we conduct the calculation in reverse: First, an effective bead diameter $\seff^\text{bead}$ is determined by applying the BH rule to the repulsive part of the bead-bead potential $U_{\text{rep}}(r_{ij})$ (\equationabbrv~(\ref{Eq:Urep})), its value is always slightly above $1.0\, \sigma$ (see \tableabbrv~\ref{tab:sigma_b2}, Appendix). Then, the 6 beads of the model constitute the occupied, ``dry'' volume $6 \cdot (\pi/6) (\seff^\text{bead})^3$ of the protein without hydration shell, which is approximately the BSA volume in protein powder form (and is nearly identical the molecule volume calculated in \refabbrv~\citenum{Hirschmann2023}). The dry volume can be converted to an effective sphere diameter $\seffdry$ of an equivalent sphere with the same volume, i.e. $\seffdry=6^{1/3}\seff^\text{bead}$. Hence, in contrast to the previous methods no $\Ueff(r)$ is computed here.

\begin{figure}[h]
    \centering
    \includegraphics[width=1\linewidth]{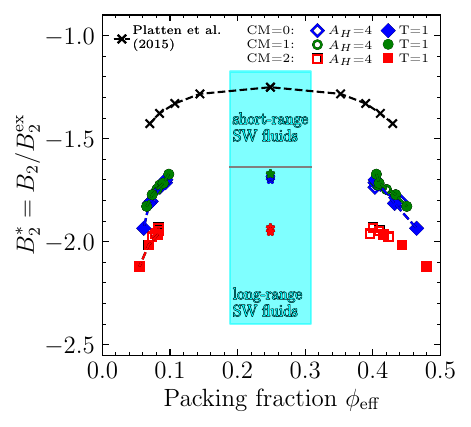}
    \caption{
    Reduced second virial coefficient $\Bts=B_2/\Btex$ with $\Btex=(2\pi/3) (\seffBINO)^3 $ as a function of the volume packing fraction $\phi=\phi_{\text{eff}}=(\pi/6) ({\seffBINO})^3\rho$ for the six different parameter combinations. The cyan shaded area represents the range of critical $\Bts$ values for short-range ($\delta\leq 0.25$) and long-range ($\delta\geq0.25$) hard sphere square well fluids calculated by Gazzillo et al..\cite{Gazzillo2013} Black crosses represent the reference hard sphere square well binodal taken from Platten et al.\cite{Platten2015} (\figureabbrv~2 and \tableabbrv~2 in that reference), which was used to obtain $\seffBINO$ as described in \sectionabbrv~\ref{sec:res_sigmas} by matching the critical points. Dashed lines are guides to the eye.
    }
    \label{fig:phasediagramb2}
\end{figure}

(iv): $\seffBINO$. Having calculated the critical molecule number densities $\rho_c$ of our systems (see \tableabbrv~\ref{tab:resultsanisotropic}), we can define an effective volume packing fraction $\phi^c_{\text{eff}}$ at the critical point via $ \phi^c_{\text{eff}}=(\pi/6) ({\seffBINO})^3\rho_c$. Subsequently, we equate $\phi^c_{\text{eff}}=\phi^c_{\text{SW}}$, where $\phi^c_{\text{SW}}=0.249$ is the well defined volume packing fraction at the critical point of a short-range, hard sphere square well fluid (taken from \refabbrv~\citenum{Platten2015}, \tableabbrv~2 in that reference and the system with reduced well width $\delta=0.05$), which is approximately representative of the short-range potentials to which the Noro-Frenkel rule applies. This analysis yields $\seffBINO\approx 1.98...2.06\,\sigma$, see \figureabbrv~\ref{fig:B_sig_combo}(b) and \tableabbrv~\ref{tab:sigma_b2} (Appendix). Furthermore, we calculate the reduced second virial coefficients for the binodal points presented in \figureabbrv~\ref{fig:phasediagram}, using $ \phi_{\text{eff}}=(\pi/6) ({\seffBINO})^3\rho$ and $\Btex=(2\pi/3) (\seffBINO)^3 $. The resulting phase diagrams in the $\phi_{\text{eff}}\text{--}\Bts$ plane are shown in \figureabbrv~\ref{fig:phasediagramb2}, along with the phase diagram of the aforementioned, short-range, reference square well system (black crosses). The packing fractions of our binodals approximately match the binodals of the reference, which is in effect similar to the construction\cite{Platten2015} used by the group of Stefan Egelhaaf mentioned in the introduction and further discussed in \sectionabbrv~\ref{sec:relToEx}. Note that if a longer ranged hard sphere square well fluid with $\delta\approx0.25$ would be used as reference, then $\phi^c_{\text{SW}}\approx0.2$,\cite{Platten2015} and therefore $\seffBINO$ would decrease by approx. $7 \%$. 

\figureabbrv~\ref{fig:phasediagramb2} also shows $B_2/\Btex$--ranges in which short-range (SR, $\delta \leq 0.25 $) and long-range (LR, $\delta \geq 0.25 $) hard sphere square well systems are critical (cyan shaded area).\cite{Gazzillo2013} This total range is nearly identical to the range of the isotropic Noro-Frenkel fluids depicted in \figureabbrv~\ref{fig:B2B2ex_betaBeta_combo}(a) (also cyan shaded area), but allows for a distinction between different ranges of attraction.

All previously discussed effective diameters are listed in \tableabbrv~\ref{tab:sigma_b2} (Appendix) and shown in \figureabbrv~\ref{fig:B_sig_combo}(b). Evidently, method (i) yields the smallest $\seff$ compared to the other methods, whereas methods (ii)-(iv) yield higher values, which are generally relatively close to each other for all investigated systems. Systems with constant $T$ roughly yield identical results as systems with constant $A_H$ for the same $\text{CM}$. Adding charges to the molecules significantly increases their effective diameter only for method (i), as the increased electrostatic repulsion at short distances is directly present in $\Ueff$, whereas methods (ii) and (iii) are (for the systems with variable temperature) only indirectly sensitive to the parameter $\text{CM}$, i.e. solely via the shifted inverse temperature $\beta_c$ which enters the BH rule. The latter is also responsible for all $\seffISO$ being smaller than their respective $\seffNF$, as $T_c^{\text{iso}}/T_c\approx2...3$. 

\begin{figure*}[!htb]
    \centering
    \includegraphics[width=1\linewidth]{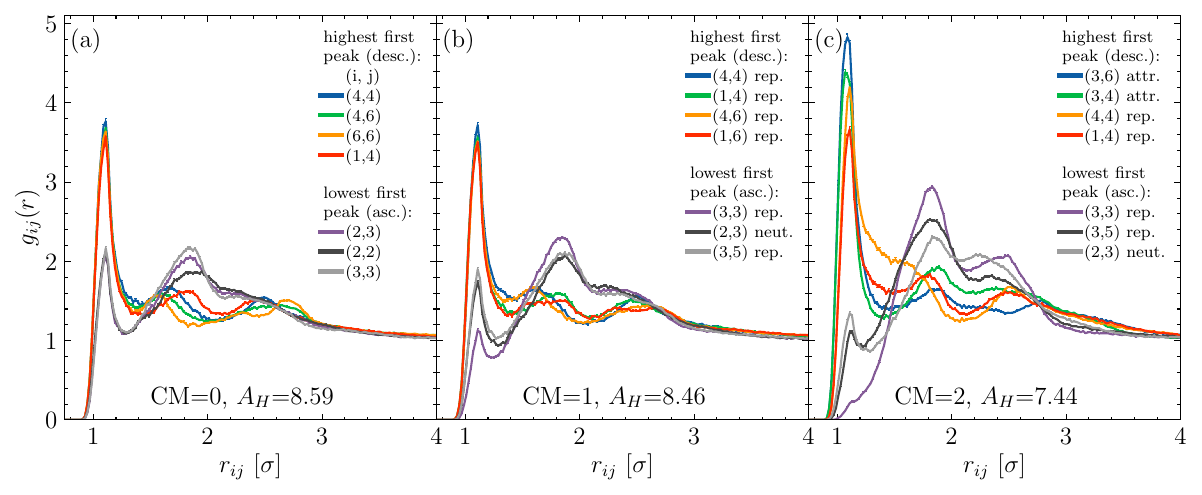}
    \caption{
    Bead-bead correlation functions $g_{ij}(r_{ij})$ with $r_{ij}$ representing the center-to-center distance of bead $i$ to bead $j$, for systems with constant temperature $T=1$ and variable $\text{CM}$ at a dilute molecule number density $\rho=5.75 \times 10^{-3}~{\sigma^{-3}}$ (for the locations in the phase diagram see the colored crosses in \figureabbrv~\ref{fig:phasediagram}(b)). Only selected $(i,j)$  combinations that exhibit the highest and lowest contact peak values are shown. Attractive, repulsive and neutral charge configurations are indicated in the legend. Error bars show the standard deviation of the mean. 
    }
    \label{fig:beadcorrelations}
\end{figure*}

Note that methods (i) and (ii) are in spirit similar to each other, as they derive the effective diameter only from the two-body interaction and the pair potential we attribute to the model. Method (iii) is based on an experimentally measured quantity, while method (iv) is based on the critical point, a collective property.

\subsection{Bead-bead correlations}\label{sec:res_beadcorr}

In order to evaluate how charge-anisotropy influences the preferred bead contacts in our simulations we compute bead-bead correlation functions $g_{ij}$ in \figureabbrv~\ref{fig:beadcorrelations} (note that $g_{ij}=g_{ji}$) for our coarse-grained molecules. We choose a relatively low density $\rho=5.75 \times 10^{-3}~{\sigma^{-3}}$ outside the coexistence region in the gas region of the phase diagram (colored crosses in \figureabbrv~\ref{fig:phasediagram}(b)) to mitigate packing effects, which would complicate a separation with effects stemming from possible patchy interactions. 

Starting from the uncharged molecule state $\text{CM}=0$ one notices first-peak contacts are most pronounced for combinations $(i,j)$ of exposed beads at the edges of the molecule, i.e. beads 1,4 and 6 and least pronounced for beads that are closer to the center of the molecule, i.e. beads 2, 3 and 5 (compare to \figureabbrv~\ref{fig:BSA_molec}). Adding the native charges to the BSA models ($\text{CM}=1$) does not introduce new $(i,j)$ combinations to the top first-peak bead-bead correlations, indicating that for this system the van der Waals and not the electrostatic interaction determines association of the molecules. This is additionally supported by the observation that only bead combinations with like charge are within the top correlation functions. However, increasing $\text{CM}$ to 2 finally switches the dominant interaction mechanism from van der Waals to electrostatics, as now the strongest charged beads exhibit either the highest first-peak correlations, i.e. (3,6) and (3,4) if oppositely charged, or the lowest first-peak correlations, i.e. (3,3) and (3,5) if like charged. Intriguingly, despite the electrostatic dominance the bead-bead combinations (4,4) and (1,4) favoured by the molecule's geometry combined with the dispersion force still exhibit the third and fourth highest correlation functions. We thus conclude that for our model there is a crossing-over between $\text{CM}=1$ and $\text{CM}=2$, where the mechanism of preferred contact selection between beads switches from van der Waals to electrostatic interactions, where the former's influence still persist to some extent. Additionally, we infer that this analysis provides arguments that our BSA model effectively exhibits three patches (within a KF picture), due to the three sites at the edges of the CG molecule (beads 1, 4 and 6), that show preferred contact propensity in $g_{ij}$.

\section{Discussion: Effective patchiness from critical points}\label{sec:discussion}
\subsection{Reduced second virial coefficient $\boldsymbol{\Bts}$ via method (i)}\label{sec:discussion_method_i}

In their seminal paper,\cite{Noro2000} Noro and Frenkel have shown that systems with isotropic, short-range attractive pair potentials can be mapped onto hard sphere square well systems (with the same $B_2$ and the effective hard sphere diameter $\seffNF$ determined with the Barker-Henderson rule), with the effect that the reduced virial coefficient $\Bts$ 
at the critical temperature is approximately constant over quite a range of pair potentials and given by $\Bts=\Bnf \approx -1.5$. The Noro-Frenkel rule does not hold for patchy particles, though. Foffi and Sciortino\cite{Foffi2007} have shown that for Kern-Frenkel (KF) particles \cite{kernfrenkel2003} (hard spheres with attractive surface patches defined via narrow cones with opening angle $\theta_0$ and with reduced attraction range $\delta_{\text{KF}}$) with $M=3$, 4, 5 patches $\Bts$ is approximately a function of $M$ only, but the actual values are very different from $\Bnf$ ($\approx -30$ for $M=3$, $\approx -4 ...-5 $ for $M=4$ and $\approx -2 ...-3 $ for $M=5$). 

\begin{figure*}[!hb]
    \centering
    \includegraphics[width=0.9\linewidth]{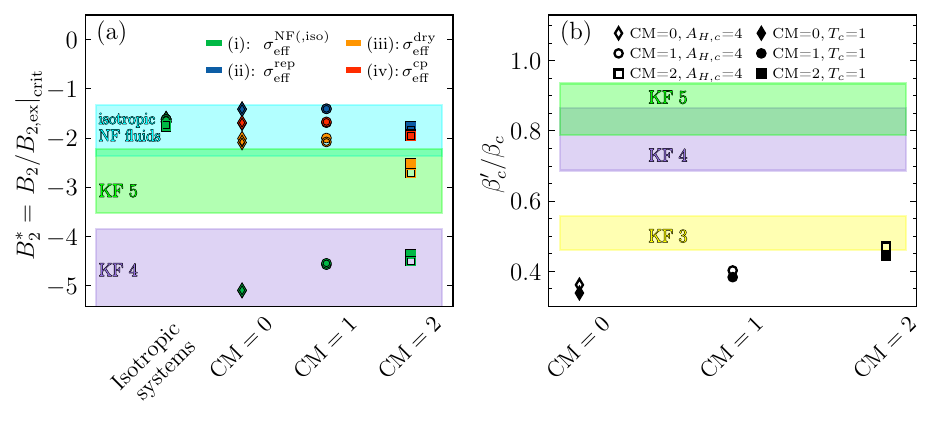}
    \caption{(a) Reduced second virial coefficients $B^*_2=B_2/\Btex$, with $\Btex=2\pi\sigma_{\text{eff}}^3/3$ and where $\sigma_{\text{eff}}$ is taken as $\sigma_{\text{eff}}^{\text{NF(,iso)}}$, $\seffrep$, $\seffdry$ or $\seffBINO$ (see \sectionabbrv~\ref{sec:res_sigmas} and \figureabbrv~\ref{fig:B_sig_combo}(b)) for the anisotropic, critical BSA systems together with $B^*_2$ values of the corresponding critical isotropic systems, as well as regions in which isotropic fluids (\tableabbrv~1 in \refabbrv~\citenum{Noro2000}, cyan shaded area) and Kern-Frenkel particles with $M=4,5$ patches are critical (\tableabbrv~1 in \refabbrv~\citenum{Foffi2007}, violet and green shaded areas). (b) Ratios $\beta_c'/\beta_c$ for all critical BSA systems together with regions in which $\beta_c'/\beta_{c,\text{KF}}$ falls for different numbers of patches $M=3,4,5$ in the Kern-Frenkel model (yellow, violet and green shaded areas). Latter regions have been calculated as explained in the main text with values taken from \refabbrvplural~\citenum{Noro2000} and \citenum{Foffi2007}. } 
    \label{fig:B2B2ex_betaBeta_combo}
\end{figure*}

In the KF model, let $\eKF$ be the attractive interaction strength of two interacting KF patches. The corresponding effective potential is a square well potential with well depth $\epsilon'$, which is related to $\eKF$ by:
\be 
 \label{eq:sweps}
 \exp( \beta \epsilon') = \chi^2 ( \exp( \beta \eKF)  -1) +1,
\ee 
where $\chi=(M/2)(1-\cos\theta_0)<1$. It is interesting to ask for the relation of critical temperatures in the orientation-dependent KF system vs. in the corresponding effective square well system. For the inverse critical temperature $\bKF$ of KF we use the data taken from \refabbrv~\citenum{Foffi2007}. The inverse critical temperature $\beta_c'$ of the effective square well system can be safely estimated using the Noro-Frenkel rule and the reduced $B_2$ in the square well system, i.e.:
\be 
 \left. B_2/\Btex \right|_\text{crit} =(1 -\exp(\beta_c' \epsilon'))( (1 + \delta)^3 -1 ) +1 \approx \Bnf,  
\ee
where $\delta$ is the range of the square well attraction in units of the hard sphere diameter. Here, $\epsilon'$ is determined from $\eKF$ from \equationabbrv~(\ref{eq:sweps}), with $\beta=\bKF$ (the inverse critical temperature of the KF system) and we set $\delta=\delta_{\text{KF}}$. From the data in \refabbrv~\citenum{Foffi2007} for various KF systems we find a ratio $\beta_c'/\bKF \approx 0.46...0.56$ ($M=3$) and $\beta_c'/\bKF \approx 0.69...0.87$ ($M=4$) and $\beta_c'/\bKF \approx 0.79...0.94$ ($M=5$). It is clear that the critical temperature of the effective system (particles interacting with $\Ueff$) is always noticeably higher than the one of the KF system, but the ratio is not only a function of the number $M$ of patches. There is a clear separation between the values for all $M=3$ systems from the rest, but there is a considerable overlap in the $\beta_c'/\bKF$-ranges for $M=4$ and 5, see \figureabbrv~\ref{fig:B2B2ex_betaBeta_combo}(b). All values are, however, significantly smaller than 1 which is the limit of isotropic potentials.

Note that in the KF analysis above the anisotropy only arose from the patchy attractions, while repulsions are isotropic hard sphere in both the KF and the effective system. This is different in our system where excluded volume interactions are very anisotropic and the effective isotropic repulsions are very soft.

These considerations can be applied to our model since $\Ueff$ and the corresponding critical temperatures in the effective systems
have been determined in \sectionabbrv~\ref{sec:res_ueff}, we refer to the results for $\seffISO$ and $\Bts$ in \figureabbrv~\ref{fig:B_sig_combo}(b) and \figureabbrv~\ref{fig:B2B2ex_betaBeta_combo}(a) (respectively) and also in \tableabbrv~\ref{tab:sigma_b2} (Appendix). For all investigated isotropic systems, $\Bts \approx -1.6...-1.7$ which is different from $\Bnf\approx -1.5$. This is due to the rather long range in the attractions of $\Ueff$ and matches very well results in Noro and Frenkel's work \cite{Noro2000} for wide square wells: $\Bts$ changes from $-1.65$ to $-2.26$ if the reduced radial range $\delta$ of the attractive square well changes from 0.5 to 1. The effective diameters $\seffISO\approx 1.2...1.4 \, \sigma$ for all systems (see \figureabbrv~\ref{fig:B_sig_combo}(b)), and it is interesting to note that these diameters are only $\approx$ 20...40\% larger than the bead diameter $\sigma$.

For the assessment of ``patchiness'', the values of $\beta_c'/\beta_c$, as well as $\Bts$ and $\seffNF$ at the critical temperature of the anisotropic system are relevant. As can be seen in \figureabbrv~\ref{fig:B2B2ex_betaBeta_combo}(b) and \tableabbrv~\ref{tab:sigma_b2} (Appendix), the values of $\beta_c'/\beta_c$ are around 0.34...0.47 with a moderate spread and all values of $\Bts$ are between $-4$ and $-5$ with a likewise moderate spread. Regarding charge dependence, $\seffNF$ exhibits a systematic increase for a rising CM value (see \figureabbrv~\ref{fig:B_sig_combo}(b), corresponding to the observation of a larger repulsive core in $\Ueff$). Hence, although the magnitude of $B_2$ is growing with increasing CM (see \figureabbrv~\ref{fig:B_sig_combo}(a)), the charge dependence of $\seffNF$ (and consequently also of $\Btex$) leads to the monotonic and moderate charge dependence in $\Bts$ (similar to $\beta_c'/\beta_c$). 

Comparing this to the KF results, these values correspond to 4 patches (from $\Bts$) or 3 patches for the $\text{CM=2}$ systems (from $\beta_c'/\beta_c$), which points to a considerable effective anisotropy in the BSA model.

\subsection{Alternative determinations of $\boldsymbol{\Btex}$ via methods (ii)-(iv) and their influence on $\boldsymbol{\Bts}$}\label{sec:discussion_method_ii_iv}

The normalization of $B_2$ with $\Btex=2\pi\seff^3/3$, the second virial coefficient for the excluded volume interaction of the equivalent sphere system, leaves a certain ambiguity in the definition of $\Btex$ which becomes manifest in a system where the repulsive interactions are explicitly anisotropic. It is interesting to explore resulting $\Bts$ if the additional $\seff$ of \sectionabbrv~\ref{sec:res_sigmas} are used.

The values for $\seffrep$, $\seffdry$ and $\seffBINO$ are around $1.9...2.1 \, \sigma $ (approximately charge-independent) and relatively close to each other, see also  \figureabbrv~\ref{fig:B_sig_combo}(b) and \tableabbrv~\ref{tab:sigma_b2} (Appendix).
All three effective diameters are considerably larger than $\seffNF$ with the immediate consequence that $\Bts$ is reduced in magnitude and they all exhibit a stronger charge dependence than before: For method (ii) (with $ \seff = \seffrep$) we find $\Bts \approx -1.4$ for the systems with $\text{CM}=0$ and 1, and $\Bts \approx -1.8$ for the systems with $\text{CM}=2$. For method (iii) (with $ \seff = \seffdry$) the difference between $\text{CM}=0,1$ on the one side and $\text{CM}=2$ on the other side can be seen as well: $\Bts \approx -2.0...-2.1$ ($\text{CM}=0,1$) and $\approx -2.5...-2.7$ ($\text{CM}=2$). Method (iv) yields approximately values that fall in the middle of (ii) and (iii): $\Bts \approx -1.7$ ($\text{CM}=0,1$) and $\Bts \approx -1.9$ ($\text{CM}=2$).  Compared with the values from the KF model, we would argue that according to methods (ii)--(iv) our BSA model is fairly isotropic for $\text{CM}=0,1$. However, the deviating values of $\Bts$ for the $\text{CM}=2$ systems are intriguing. Here method (iii) yields results compatible with a 5 patch KF model.

The ambiguity in defining $\Btex$ drops out if we form the ratio of critical $B_2$-values
\be \label{eq:bratio}
   r =  \frac{ B_2(\text{CM}=2)}{ B_2 (\text{CM}=0)}
\ee
which yields $\approx 1.24...1.29$. Using the results for different systems in \tableabbrv~1 of \refabbrv~\citenum{Foffi2007}, we estimate for the KF model $r\approx 4.12...7.35$ for the ratio between a 3-patch and a 4-patch model, and $r \approx 1.09...2.56$ for the ratio between a 4-patch and a 5-patch model. The latter range includes our results, such that one could conclude that our BSA model is moderately patchy, around 4 patches for $\text{CM}=2$ and 5 patches for $\text{CM}=0$.

As a remark, we note that the term ``patchiness'' used in this study refers solely the kind present in the KF model, i.e. a purely attractive one. There are other kinds of interesting model systems without uniform van der Waals attraction exhibiting patches that repel each other, for example inverse patchy colloids\cite{Bianchi2011} (IPCs), lattice charge models\cite{Kohara2022} or continuous charge patchy models\cite{Bozic2018}. Such repulsive patches are obviously also present in our CG model, due to same-charge interactions, and they are overall less dominant than the opposite-charge interactions (otherwise $T_c$ would not increase by increasing $\text{CM}$). By the choice of our KF/NF reference systems the effects of these two types of patches are mixed to an effective, attractive patchiness. Here, a systematic study that separates these effects would be intriguing, however, no law of corresponding states that would enable a investigation similar to this one has been observed for IPCs so far.\cite{Notarmuzi2024}

\subsection{Critical density $\boldsymbol{{\rho}_c}$} \label{sec:critDens}
\begin{figure}[]
    \centering
    \includegraphics[width=1\linewidth]{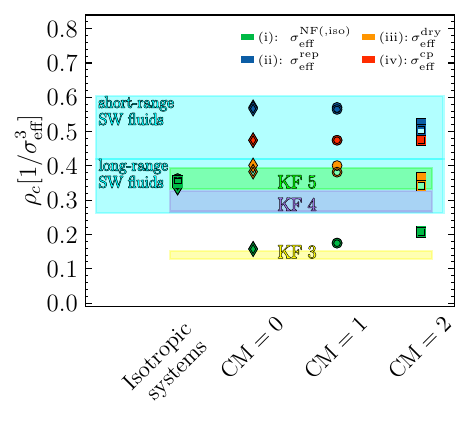}
    \caption{
    Critical densities $\rho_c$ (\tableabbrv~\ref{tab:resultsanisotropic}) in units of $1/\seff^3$, where $\seff$ is taken as $\sigma_{\text{eff}}^{\text{NF(,iso)}}$, $\seffrep$, $\seffdry$ or $\seffBINO$ (see \sectionabbrv~\ref{sec:res_sigmas} and \figureabbrv~\ref{fig:B_sig_combo}(b)). The shaded areas correspond to the critical densities for KF particles reported in \refabbrv~\citenum{Foffi2007} and to short- ($\delta \leq 0.25 $) and long-range ($\delta \geq 0.25 $) hard sphere square well systems reported in \refabbrv~\citenum{Platten2015}.
}
    \label{fig:rhoc}
\end{figure}

Also the critical density ${\rho}_c$ is sensitive to the number of patches within the KF model, but in contrast to $\Bts$ the range of critical densities for long-range isotropic fluids overlaps with the corresponding range for KF fluids with 4 and 5 patches (see \figureabbrv~\ref{fig:rhoc}). The corresponding quantity in our model systems is the critical density (\tableabbrv~\ref{tab:resultsanisotropic}) evaluated in units of $1/\seff^3$, using the various definitions from \sectionabbrv~\ref{sec:discussion_method_i} and \ref{sec:discussion_method_ii_iv} (\figureabbrv~\ref{fig:B_sig_combo}(b) and \tableabbrv~\ref{tab:sigma_b2} (Appendix)), see the symbols in \figureabbrv~\ref{fig:rhoc}. The analysis confirms the conclusions from \sectionabbrv~\ref{sec:discussion_method_i} and \ref{sec:discussion_method_ii_iv} from the analysis of $\Bts$, especially regarding the difference between $\seffNF$ on the one hand and the alternative definitions (ii)-(iv) from \sectionabbrv~\ref{sec:discussion_method_ii_iv}. If the latter are used, the critical densities for the $\text{CM}=0$ systems are in the range of isotropic short-range square well fluids and decrease slightly with increasing the charge. Only for $\seffdry$, the highly charged $\text{CM}=2$ systems appear slightly patchy (5 KF patches), however, overlapping with the range of isotropic long-range square well fluids. 
Overall, this is comparable to the slight decrease of $\Bts$ with increasing charge seen in \figureabbrv~\ref{fig:B2B2ex_betaBeta_combo}(a) for the alternative definitions of $\seff$. The relative decrease of ${\rho}_c(\text{CM}=2)$ vs. ${\rho}_c(\text{CM}=0)$ of about $\SI{10}{\percent}$ roughly corresponds to the relative decrease in ${\rho}_c$ of KF 5-patch to 4-patch models (and would constitute an estimate  independent of $\seff$). If however method (i) is used for the anisotropic systems, the critical densities for the $\text{CM}=0$ systems are compatible with a 3-patch KF system and the charges tend to increase the critical density towards the one of a 4-patch KF system. This effect of charge is similar to the increase of $\Bts$ for this definition of the effective diameter, see \figureabbrv~\ref{fig:B2B2ex_betaBeta_combo}(a), the only difference being that according to the critical density analysis the CG model has an overall more pronounced patchy nature. The critical densities of the isotropic systems are in the range of isotropic long-range fluids (in agreement with the long-range character of $U_{\text{eff}}$ in \figureabbrv~\ref{fig:ueff}).

\subsection{Relation to experiment}\label{sec:relToEx}

Scattering experiments on BSA solutions in monovalent salt indicated stability of the BSA solution.\cite{Zhang2007} Available experimental data for $B_2$ in sodium chloride solution at pH=7.4 and ambient temperatures were fitted in \refabbrv~\citenum{Kozlowska2021} using a 20-bead model with DLVO forces, similar to ours. The resulting $B_2$ values (in our unit $\sigma^3$) in the range $-5...25$ are considerably larger than the $B_2$ values at the critical point for our models (which are in the range $-28...-37$, see Fig.~\ref{fig:B_sig_combo}(a)). Since the fit of \refabbrv~\citenum{Kozlowska2021} was for high salt concentrations and therefore effective charge screening, one should compare to the maximal value for the chargeless model ($CM=0$), $B_2\approx-28$. Thus one would conclude that a hypothetical critical temperature in a sodium chloride solution is way below the freezing point of water which is outside of physical interest. 

In contrast to sodium chloride solutions, liquid-liquid phase separations have been observed in BSA solutions with trivalent cations.\cite{Zhang2008} The necessary attraction has been proposed to be due to cation bridges at specific places on the BSA surface, i.e. patchy attractions in essence.\cite{RoosenRunge2014} The phase separation occurs near the point of zero charge. Measurements using small angle X-ray scattering (SAXS) of $B_2/\Btex$ \cite{Braun2017,Matsarskaia2018,Surfaro2023,Braun2018} have indicated though that $\Bts \approx -1.5...-1.8$, which is of the order of $\Bnf\approx-1.5$, and thus one would ascribe no patchiness to BSA. However, in these works the effective hard sphere diameter $\seff$ used in $\Btex$ was fitted from an ellipsoidal form factor and gave $\seff \approx \SI{9.2}{\nm} $, which is larger than our $\seff$ by a factor $1.5...2.6$. This particular ellipsoidal form factor fit is probably due to clusters. Note that a similar form factor fit in the SAXS measurements of \refabbrv~\citenum{Zhang2007} (with monovalent salt, less attraction and thus less propensity for cluster formation) gave $\seff\approx \SI{6.7}{\nm}$ which is much closer to our $\seffrep\approx \SI{6.3}{\nm} $ which can be considered similar in spirit. Thus, using our range of $\seff$, $\Bts$ in the BSA-trivalent cation system is estimated to be around $-4.6...-32$ which indicates much more patchiness and also a larger patchiness than found in our model (ranging between $-1.4...-5.1$). Our model has not been designed with very specific bridging attractions, rather we have put the effect of the attraction into an increase of $A_H$ and the attraction between positive and negative charges at the BSA surface. We would expect that $\Bts$ attains larger negative values if specific bridging sites are included in our model. 

A somewhat different perspective on the validity of the Noro-Frenkel rule for protein solutions has been given in \refabbrv~\citenum{Platten2015}, discussing the phase diagrams of globular proteins, mainly of lysozyme. The determination of a suitable effective diameter and thus of $\Bts$ uses the liquid-liquid binodal, which is in effect similar to method (iv) of Sec.~\ref{sec:res_sigmas}. Specifically, the argument in \refabbrv~\citenum{Platten2015} concerning the validity of the Noro-Frenkel rule proceeds as follows: From an extended body of simulation data for binodals of isotropic fluids a ``Noro-Frenkel band'' of binodals is defined in the $\phi\text{--}\Bts$ plane (packing fraction vs. reduced second virial coefficient). Considering now a protein solution, an effective diameter is defined by fitting the gas part of its binodal to the binodal of a square well fluid with reduced well width $\delta=0.05$ (this binodal is located approximately in the middle of the ``Noro-Frenkel band''). As a result, the investigated lysozyme binodals nicely fall into this band and $\Bts \approx \Bnf$. Thus, \refabbrv~\citenum{Platten2015} argues, that the Noro-Frenkel rule is valid and would presumably also hold for other globular proteins, i.e. from critical liquid-liquid demixing no patchiness can be inferred. As a result, a such defined diameter is considerably larger than values derived from an effective, isotropic DLVO potential (as also seen in \refabbrv~\citenum{Platten2015}). This is in agreement with our findings for $\seffBINO \approx 2.0$ (whose derivation is in effect similar to the construction of \refabbrv~\citenum{Platten2015}), which is larger than $\sigma_{\text{eff}}^{\text{NF(,iso)}} \approx1.2...1.6$ (from the orientation-averaged effective potential). Consequently, the critical points of all systems with varying $\text{CM}$ are well within the region where isotropic fluids are critical, see \figureabbrv~\ref{fig:phasediagramb2} (Appendix). If using a finer distinction of $\Bts(T_c)$ values, that discriminates short and long-range hard sphere square well systems, the $\text{CM}=2$ system clearly corresponds to longer ranged wells ($\delta\geq0.25$), whereas the $\text{CM}=0,1$ systems fall in an intermediate region where $\delta\approx0.25$.

\section{Conclusion}\label{sec:conclusion}

We have investigated the phase diagram and critical points of an anisotropic coarse-grained model for the globular protein BSA. Besides the shape anisotropy through the spatial arrangement of the beads, we investigated also the dependence of the critical point on the charge asymmetry, realized by different bead charges at the isoelectric point. We have attempted to quantify the anisotropy effects on the critical point through an ``effective patchiness'' which is built on the suggestion by Foffi and Sciortino \cite{Foffi2007} for an extended Noro-Frenkel rule \cite{Noro2000} for patchy particles. A difficulty arises from the fact that in the analysis of Foffi and Sciortino the repulsive core of the patchy particles are isotropic hard spheres whereas in our case the repulsive core is soft and anisotropic.

We have argued that the analysis of the effective, orientation-averaged pair potential $\Ueff$ provides a useful link: first, it allows to define an effective hard sphere diameter $\seffNF$ and a repulsive part $\Btex$ of the second virial coefficient and secondly, the ratio of critical temperatures of an isotropic model interacting with $\Ueff$ and of the fully anisotropic model contains a definite patch dependence. The reduced second virial coefficient at the critical point $\Bts = B_2/\Btex(T_c)$ points to an effective patchiness of 4 Kern-Frenkel patches ($\Bts \approx -4 ... -5$), whereas the ratio of critical temperatures points to a 3-patch model for $\text{CM}=2$, see \figureabbrv~\ref{fig:B2B2ex_betaBeta_combo}. Through computing the ratio $r$ of $B_2$ (\equationabbrv~(\ref{eq:bratio})), the dependence on $\Btex$ drops out and for our BSA model we see that the $\text{CM}=0$ and $\text{CM}=2$ charges lead to an effective anisotropy of approximately 4 ($\text{CM}=2$) to 5 ($\text{CM}=0$) Kern-Frenkel patches.

Through its dependence on $\Btex$, the $\Bts$-criterion does not appear to be unique. We have investigated two other sensible criteria for defining $\Btex$ and $\seff$ directly related to the molecular properties (through the repulsive part of the bead-bead potentials and the dry protein volume, yielding $\seffrep$ and $\seffdry$) and obtain larger effective diameters and consequently also values $\Bts \approx -2$ (and larger) which are mostly compatible with the Noro-Frenkel rule for isotropic particles, except for $\seff=\seffdry$ and $\text{CM}=2$, see \figureabbrv~\ref{fig:B2B2ex_betaBeta_combo}(a). 

The analysis presented here shows that care is needed in the definition of an effective hard sphere diameter when working with a $\Bts$-criterion. This has become evident when (as a third alternative) applying an analysis similar to the one by Platten et al.\cite{Platten2015} on the phase behavior of lysozyme, yielding $\seffBINO$. Here an effective hard sphere diameter was defined through matching the critical points of the proteins to a square well system, which again resulted in an approximate validity of the Noro-Frenkel rule.

It appears difficult to straightforwardly apply rules for isotropic systems with reasonably well defined excluded volume to anisotropic systems. Our work is an example how much the results for $\Bts$ can vary when applying different methods which all appear reasonable. In this respect, computing the isotropic potential $U_{\text{eff}}$ offers two advantages. First, the definition of a hard sphere diameter via Barker-Henderson integration (as applied for soft isotropic systems in \refabbrv~\citenum{Noro2000}) and second the possibility of using the ratio of critical temperatures between the anisotropic and the effective isotropic system as a variable to infer patchiness.

\section*{Author Contributions}

Jens Weimar and Frank Hirschmann: Data curation, Formal Analysis, Investigation, Methodology, Software, Visualization, Writing – original draft, Writing – review \& editing. Martin Oettel: Conceptualization, Funding acquisition, Methodology, Project administration, Resources, Supervision, Validation, Writing – original draft, Writing – review \& editing

\section*{Conflicts of interest}
There are no conflicts to declare.

\section*{Acknowledgements}
The authors acknowledge support by the state of Baden-W\"urttemberg through bwHPC and thank Dr. Fajun Zhang for insightful discussions on this manuscript and for providing experimental structure factors.

\section*{Appendix}

For completeness, we show in \tableabbrv~\ref{tab:sigma_b2} all investigated observables of the six systems at their respective critical points.

\begin{table*}[ht]
    \centering
    \caption{Summary of critical parameters and effective molecule diameters $\seff$ (in units of $\sigma$) at criticality calculated via the different methods (i)-(iv) presented in \sectionabbrv~\ref{sec:res_sigmas} for the anisotropic and isotropic systems. We estimate the total relative error for $\seff$ and $\Bts$ to be $ \pm \, \SI{0.7}{\percent}$ and $ \pm \, \SI{2.6}{\percent}$ (respectively) for methods (i) and (ii),  $ \pm \, \SI{0.9}{\percent}$ and $\pm \, \SI{3.4}{\percent}$ for method (iii) and $ \pm \, \SI{0.4}{\percent}$ and $\pm \, \SI{1.7}{\percent} $ for method (iv)   }  
    \begin{tabular}{@{\extracolsep{\fill}}lllllllllllllllll}
        \hline
        & & \multicolumn{11}{c}{Anisotropic systems} & \multicolumn{4}{c}{Isotropic systems} 
        \\\cmidrule[0.02cm](lr){3-13} \cmidrule[0.02cm](lr){14-17} 
        & & \multicolumn{2}{c}{} & \multicolumn{2}{c}{ (i): $\seffNF $ } & \multicolumn{2}{c}{ (ii): $\seffrep$ } & \multicolumn{3}{c}{ (iii): $\seffdry$ } & \multicolumn{2}{c}{ (iv): $\seffBINO$ } & \multicolumn{2}{c}{{  }} & \multicolumn{2}{c}{{ (i):  $\seffISO$ }} 
        \\\cmidrule[0.02cm](lr){5-6} \cmidrule[0.02cm](lr){7-8}  \cmidrule[0.02cm](lr){9-11} \cmidrule[0.02cm](lr){12-13} \cmidrule[0.02cm](lr){16-17}
        System & CM & $T_c$ & $A_{H,c}$ & $\Bts$ & $\seffNF$ & $\Bts$ & $\seffrep$ & $\Bts$ & $\seffdry$ & $\seff^\text{bead}$ & $\Bts$ & $\seffBINO$ & $T_c^\text{iso}$ & $\nicefrac{\beta_c'}{\beta_c}$ & $\Bts$ & $\seffISO$ \vspace{0.02cm}
        \\  \hline
		I & 0 & 0.405 & 4 & -5.10 & 1.39 & -1.41 & 2.14 & -2.01 & 1.90 & 1.05 & -1.70 & 2.01 & 1.12 & 0.36 & -1.66 & 1.23 \\
		II & 0 & 1 & 8.59 & -5.09 & 1.37 & -1.41 & 2.10 & -2.09 & 1.85 & 1.02 & -1.69 & 1.98 & 2.95 & 0.34 & -1.61 & 1.19 \\
		III & 1 & 0.411 & 4 & -4.57 & 1.44 & -1.40 & 2.14 & -2.00 & 1.90 & 1.04 & -1.69 & 2.01 & 1.02 & 0.40 & -1.67 & 1.28 \\
		IV & 1 & 1 & 8.46 & -4.54 & 1.42 & -1.41 & 2.10 & -2.08 & 1.85 & 1.02 & -1.67 & 1.98 & 2.61 & 0.38 & -1.62 & 1.24 \\
		V & 2 & 0.470 & 4 & -4.36 & 1.57 & -1.75 & 2.13 & -2.51 & 1.89 & 1.04 & -1.93 & 2.06 & 1.00 & 0.47 & -1.76 & 1.43 \\
		VI & 2 & 1 & 7.44 & -4.50 & 1.56 & -1.83 & 2.10 & -2.71 & 1.85 & 1.02 & -1.95 & 2.06 & 2.25 & 0.44 & -1.73 & 1.40 \\\hline
    \end{tabular}    
    \label{tab:sigma_b2}
\end{table*}

\balance
\bibliography{rsc} 
\bibliographystyle{rsc} 
\end{document}